\documentstyle[12pt,fleqn]{article}
\input psfig.tex
\begin{document}
\baselineskip=7.8mm

\begin{center}
{\Large\bf   The Evolution of Magnetic Field and
Spin Period of Accreting Neutron Stars}
\end{center}

\vspace{10.0mm}
\begin{center}

{\bf K.S. Cheng$^{1}$ and C.M. Zhang$^{1,2,3}$}

{\em 1.\,\, Department of Physics,\\ 
The University of Hong Kong,\\ 
Pokfulam Road, Hong Kong, P.R. China\\}

{\em 2.\,\, Department of Physics,\\ 
Hebei University of Technology,\\ 
Tianjin-300130, P.R. China\\}

{\em 3.Instituto de F\'{\i}sica Te\'orica\\
Universidade Estadual Paulista\\
Rua Pamplona 145\\
01405-900\, S\~ao Paulo \\
Brazil}

\end{center}

Received: \underline{\em{    } }   \\

Accepted:\underline{\em{                    } }\\

\newpage
\newcommand{\be}{\begin{equation}}
\newcommand{\ee}{\end{equation}}

\parindent=10mm
\vspace{10.0mm}

\begin{center}
ABSTRACT
\end{center}

Based on the accretion induced magnetic a field decay model, in which
a frozen field and an incompressible fluid are assumed, we obtain the
following results. (1) An analytic relation between
the magnetic field and  spin period, if
 the fastness parameter of the accretion disk is neglected.
 The evolutionary
 tracks of accreting neutron stars in the P-B diagram in our model
 are different from
 the equilibrium period lines when the influence of the fastness parameter is taken into account.
(2) The theoretical minimum spin period of an accreting neutron star 
is $\max \left(1.1{\rm ms}\left(\frac{\Delta M}{M_{\odot}}\right)^{-1}
R^{-5/14}_6 I_{45}\left(\frac{M}{M_{\odot}}\right)^{-1/2},
1.1{\rm ms}
\left(\frac{M}{M_{\odot}}\right)^{-1/2} R^{17/14}_6 \right)$, independent of 
the accretion rate (X-ray luminosity) but dependent on the total accretion mass
$\Delta M$. However, the minimum magnetic field depends on the accretion rate.
(3) The magnetic field strength decreases faster with time than the period.

{\bf  Subject headings:} stars: neutron -- stars: rotation -- stars: magnetic field
 --X-rays: stars--stars: binaries: close

\noindent

\newpage
\begin{center}
{\bf 1. Introduction}
\end{center}
The magnetic field of a neutron star has long been a complex issue and 
one which is not yet to be solved(Bhattachaya \& van den Heuvel 1991; 
Chanmugam 1992; Phinney \& Kulkarni 1994).  On the evolution of the magnetic 
field of a neutron star,  
there is not yet a commonly accepted model 
(for a general review cf. Bhattacharya and Srinivasan 1995).
The currently popular idea seems to ascribe the field decay of a neutron
star in an X-ray binary to the period  during which accretion occurs.
There is  evidence that 
the  magnetic fields of X-ray  neutron stars and recycled pulsars  
are correlated with the duration of 
the mass accretion phase, or the total amount of matter accreted(Taam \&
van den Heuvel 1986;  van den Heuvel et al 1986).                                            
In fact, Taam and van den Heuvel have already discovered a possible inverse 
correlation between the magnetic field and the estimated total mass of 
accreted matter for the binary X-ray sources. Later, Shibazaki 
et al. (1989) presented an assumed formula relating the decay of magnetic field 
with accretion  mass,which seems to reproduce the observed field-period 
relations of the recycled pulsars quite well.

Theoretically,  for explaining the accretion induced field decay, 
some suggestions and 
models have been proposed(e.g. Bisnovatyi-Kogan \& Komberg 1974; Romani 1990; 
Ruderman 1991a,b,c; Ding et al. 1993; Zhang et al. 1994; 
Urpin \& Geppert 1995; Geppert et al 1996; 
Urpin \& Konenkov 1997; Zhang et al 1997; Cheng \& Dai 1997; Zhang 1998; Ruderman, Zhu \& Chen 1998). Recently, van den 
Heuvel \& Bitzaraki(1995a, 1995b), from the statistical analysis of 24 
binary radio pulsars with nearly circular orbits and low mass companions,
discovered a clear correlation between spin period and orbital period, as
well as between the magnetic field and orbital period. These relations strongly 
suggest that an increase in the amount of accreted mass leads to a decay of the 
magnetic field, and a 'bottom' field strength of about 10$^8$ G is also
implied. 
White \& Zhang(1997) discovered that the spin periods
of LMXBs, implied by killo-hertz X-ray QPO, constitute a homogenouse group with 
spin period of about 2 milliseconds, which has little correlation with X-ray 
luminosity.
The above two recent observational statistics
seem to place some constraints on the construction of a theoretical model of 
accretion induced magnetic field decay.

In this paper, we study the evolution of a magnetic field and
spin period of accreting neutron stars
according to an accretion induced magnetic field decay
model(Zhang et al 1997; Cheng \& Zhang 1998), based on
the idea of  van den Heuvel \& Bitzaraki(1995a, 1995b) and Romani(1990)
for the mechanism of accretion induced field decay. 
 In this model, the accretion matter starts to be channeled
  onto the two polar caps  by the strong
magnetic field near the Alfven radius.
Part of the accreted matter flowing towards the equator
pushes the field lines aside
and thus dilutes the polar field strength.
The bottom field should be reached when the polar cap extends over the entire
stellar surface, which corresponds to the Alfven radius
matching the star's radius, and gives a stellar
magnetic field of about 10$^8$ G. Since the spin period of an accreting X-ray 
neutron star depends sensitively on the magnetic field
together with the influence of the fastness
parameter, some interesting results can be obtained, which are consistent with 
the recent observational data on low mass X-ray neutron stars
 by White \& Zhang (1997) and on the millisecond pulsars by 
van den Heuvel \& Bitzaraki(1995a, 1995b).

\begin{center}
{\bf 2. Models}
\end{center}
Under the assumption that the magnetic field lines of the
neutron star are frozen in
the entire crust which has homogenous average mass density, we
(Cheng \& Zhang 1998) have obtained
an analytical expression for the field evolution as follows,

\be
B = \frac{B_f}{\{1 - C {\rm exp}(-\frac{\Delta M}{M_{cr}})\}^{7/4}}
\ee
where $C = 1-x_0^2$ and $\, x_0 = (\frac{B_f}{B_0})^{2/7}$, $B_0$ is the initial magnetic field strength, $M_{cr}$ is the  crustal mass, 
$\Delta M = \dot{M} t$ and $B_f$ is the magnetic field defined by the Alfven 
radius matching the radius of neutron star, i.e., $R_A (B_f)=R$, which gives,

\be
B_f = 4.3 \times 10^8 (\frac{\dot{M}}{\dot{M}_{Ed}})^{1/2} 
(\frac{M}{M_\odot})^{1/4}R^{-5/4}_6 \,\,\,G,
\ee
where $\dot{M}_{Ed}= 10^{18} R_6 {\rm g\, s}^{-1}$ is the Eddington accretion rate and $R_6$ is the radius 
of the neutron star in units of $10^6$ cm. 

Three consistent observational conclusions can be obtained,  
(1) the field  decay is inversely related to the accreted mass, (2) the bottom 
field strength is about $10^{8}$ Gauss, and (3)  the bottom field strength is 
proportionally related to the X-ray luminosity. On the basis of the above 
solution, we will study the spin period evolution of accreting neutron stars.

To acquire the magnetic field versus period relation, we use the formula for 
the variation of the rotation due to accretion given by Ghosh \& Lamb
(1979, hereafter GL)
\be
-\dot{P} = 5.0 \times 10^{-5}[(\frac{M}{M_\odot})^{-3/7} R^{12/7}_6 
I_{45}^{-1}] B^{2/7}_{12} (PL^{3/7}_{37})^2 n(\omega_{s})\,\,{\rm s} \,{\rm yr}^{-1}\,,
\ee
where $B_{12} $ is the surface field in units of $10^{12}$ G, $I_{45}$ is 
the moment of inertia in units of $10^{45} {\rm g\, cm^2}$, $L_{37}$ is the X-ray brightness 
in units of $10^{37} {\rm erg\, s}^{-1}$ and $n(\omega_{s})$ is a dimensionless function that 
depends primarily on the fastness parameter,  

\be
\omega_{s} = 1.35[(\frac{M}{M_\odot})^{-2/7}R^{15/7}_6  ] B^{6/7}_{12} P^{-1}
L^{-3/7}_{37} \,.
\ee
For a star rotating slowly in the same sense as the disk flow 
$(\omega_{s} \ll 1 )$, GL found that $n(\omega_{s} )\sim 1.4$.
They also found
that the dimensionless function $n(\omega_{s})$ decreases with
increasing $\omega_{s}$
and becoming negative  for $\omega_{s}> \omega_{c}$. 
A simple expression for $n(\omega_{s})$ that agrees approximately with numerical 
results over the whole range of $\omega_{s}$ is, 

\be
n(\omega_{s}) = 
1.4\times \left( \frac{1-\omega_{s}/\omega_{c}}{1-\omega_{s}} \right).
\ee
GL found $\omega_{c}\sim 0.35$ from their model, but stressed  that the 
actual value of this critical fastness parameter 
was relatively uncertain. Subsequent work(Ghosh \& Lamb 1991) 
indicates that $\omega_{c}$ is unlikely to be less than 0.2, but it could  
be as large as 0.9. In the following subsections, we study the influence of the 
field decay  on the evolution of the spin period of the neutron star in an X-ray 
binary. 

\begin{center}
{\bf 2.1 Spin Evolution with a Constant Fastness Parameter}
\end{center}

If the variation of the fastness parameter can be ignored,
i.e. $n(\omega_{s})$ =1, we can solve equation (3) analytically.
First we can rewrite equation (3) as
$\frac{dP}{P^2} = - 5.0 \times 10^{-5}[(\frac{M}{M_\odot})^{-3/7} R^{12/7}_6 
I_{45}^{-1}] B^{2/7}_{f12} L^{6/7}_{37} z^{2/7} dt\,{\rm s \,  yr^{-1}}$,
where $ B_{f12} = B_f/10^{12}G$ and $z=B/B_f$. From equation (1), we obtain
$dt = - \frac{4M_{cr}}{7\dot{M}z}\frac{dz}{z^{4/7}-1}$, which gives  
$\frac{dP}{P^2} = 10^{-4}\frac{M_{cr}}{\dot{M}}(\frac{M}{M_\odot})^{-3/7}
R^{12/7}_6
I_{45}^{-1}$ $ B^{2/7}_{f12} L^{6/7}_{37}
\frac{dy}{y^2-1}\,{\rm s \,  yr^{-1}}$,
where $y=z^{2/7}$. Using equation (2) to eliminate $B_f$,
assuming mass and radius as constants and
the initial period condition $P_0 = \infty$ 
, we obtain the field-period relation (B-P) in the following analytic form,

\be
P = \frac{1.5 \,{\rm ms} \,}
{\rm{atanh(x)-atanh(x_0)}}
\, (M/M_{\odot})^{-1/2} R^{-5/14}_6 I_{45}(M_{cr}/0.1M_{\odot})^{-1}
\ee
where
$x= y^{-1} = (\frac{B_f}{B})^{2/7} \,\,, x_0=(\frac{B_f}{B_0})^{2/7}$ and 
$\rm atanh(x)=[ln(1+x)-ln(1-x)]/2$. There are two interesting limits
for equations (1) and (6). For $\Delta M \ll M_{cr}$, equations (1) and (6) can be approximated as

\be
B \approx B_f \frac{M_{cr}}{\Delta M} \propto t^{-1}
\ee
and
\be
P \approx \frac{1.5 \, {\rm ms} \,}
{x}
\, \left(\frac{M}{M_{\odot}}\right)^{-1/2} R^{-5/14}_6 I_{45}\left(\frac{M_{cr}}{0.1M_{\odot}}\right)^{-1} \propto t^{-2/7}.
\ee
For $\Delta M \gg M_{cr}$, equations (1) and (6) can be approximated as 

\be
B \approx B_f \left(1 + \frac{7}{4} {\rm exp}\left(\frac{-\Delta M}{M_{cr}}\right) \right)
\ee
and
\be
P \approx 1.1{\rm ms}
\, \left(\frac{\Delta M}{M_{\odot}}\right)^{-1} R^{-5/14}_6
I_{45}\left(\frac{M}{M_{\odot}}\right)^{-1/2}.
\ee
We should note that the period of the neutron star cannot be shorter than the
equilibrium spin-up line which represents the minimum period (Bhattacharya \& van den Heuvel 1991)
\be
P_{eq} = 2.4{\rm ms}B_9\left(\frac{\dot{M}}{\dot{M}_{Ed}}\right)^{-3/7} 
\, \left(\frac{M}{M_{\odot}}\right)^{-5/7} R^{16/7}_6, 
\ee
to which such a spin-up may proceed at the Eddington accretion rate. If we 
substitute the minimum B field in equation (11), we obtain the minimum
equilibrium period,
\be
P_{min}^{eq} = 1.1{\rm ms}
\, \left(\frac{M}{M_{\odot}}\right)^{-1/2} R^{17/14}_6. 
\ee
However, we want to point out that equation (12)
is valid only in cases where  the neutron star has accreted a sufficient amount of
matter to spin-up to that period. For a given
amount of accreted matter, the minimum period of the neutron star is 

\be
P_{min} = \max \left(1.1\left(\frac{\Delta M}{M_{\odot}}\right)^{-1}
R^{-5/14}_6 I_{45}\left(\frac{M}{M_{\odot}}\right)^{-1/2},
1.1\left(\frac{M}{M_{\odot}}\right)^{-1/2} R^{17/14}_6\right){\rm ms},
\ee
an expression which is independent of 
the accretion rate (X-ray luminosity) but depends on the total accretion
mass $\Delta M$.

We plot the field-period relation in figure 1, which 
shows the evolutionary track curves of the magnetic field and spin period
(curve 1 and curve 3).
Solid diamonds in figure 1 are observed data summarized in table 1.  
Initially, a small amount of  mass  is transferred,
 and the neutron star is spun-up from the death valley
 (Chen \& Ruderman 1993) where it has a long 
period, which causes a modest field decay and produces systems such 
as PSR0655+64 and PSR1913+16. The binaries with longer-lived accretion phases,
e.g.LMXB, will accept sufficient mass from their companions, and yield a 
substantial field decay as in the case of millisecond pulsars such as PSR1953+10 and PSR1620+21. 
Our analytic model B-P curves, which are obtained by 
assuming $n(\omega_{s})$ = 1, can go beyond the 
spin-up line (the equilibrium period line). This results from the fact that 
the influence of the fastness parameter is neglected, therefore the spin-up 
torque still exists even when the spin of the
neutron star is faster than the Keplerian angular velocity of
the accretion disk at the inner radius.
>From the analytical P-B relation, we find that the minimum period can be obtained
if the magnetic field arrives the bottom field strength(cf. equation 6). The maximum mass accreted
in LMXB from the companion could be $\sim 1.0 M_{\odot}$
(van den Heuvel \& Bitzaraki 1995a,1995b), the mass of the neutron star
after accretion could reach $\sim 2.4 M_{\odot}$ with a radius 
$R_6 \sim 1$ for realistic equations of state
(cf. Table I of Cheng \& Dai 1997) and I$_{45} \propto$ M.
So the minimum spin period given by equation (10)
is about $P_{min} \sim 1.7$ ms. Further, the more
 interesting thing is that, unlike the bottom field, the minimum
 period is  independent of the accretion rate(X-ray luminosity)
(cf. equations (10) and (12)).
 Our expression seems
to be supported by the recent work of White \& Zhang(1997). They find that 
the luminosities in the 10 samples of QPO LMXBs vary by two order of
magnitude from
$L_{36}=1$ to $L_{38}=1$, but the spin periods of the sources  diffuse into a 
narrow region from 2.76 ms to 3.8 ms. 

\begin{center}
{\bf 2.2 Spin Evolution with a Non-constant Fastness Parameter}
\end{center}

Numerical solutions (curve 2 and curve 4) for equations (1), (3), (4)  and (5) 
where the critical fastness parameter is set at 0.9
are plotted in figure 1. However,
we find that the 
influence of the fastness  parameter has little effect in the 
low magnetic field region.
The main effect of the fastness parameter is to force the evolution curves
back/below the equilibrium period line. The evolutionary curves in
the P-B diagram, however, are not sensitive to the
fastness parameter near the bottom field and/or near the minimum period. 
>From figure 1 these curves show that the field decay time scale is longer 
than the spin-up time scale because the field reaches the bottom value first, then 
the neutron star evolves towards the minimum spin period horizontally in the P-B 
diagram. The numerical solution are consistent with our analytic expressions
in equations (7) and (8).

However, the fastness parameter effect is important when the evolutionary track is close 
to the equilibrium period line, 
which means 
that the spin angular velocity of the star matches the Keplerian angular velocity 
at the inner edge of the accretion disk, and the accretion torque produced by magnetic lines immersed in the 
accretion disk tends to produce the negative(spin-down) torque.  This effect
ensures that the evolution track cannot go beyond the equilibrium period line.
It is interesting to note that none of the model B-P curves   
go along the equilibrium period line, 
which may seem confusing. However, in fact, the equilibrium 
period line only represents the final position of the evolution track in the
P-B diagram. Therefore it is not surprising that the
real evolutionary track deviates from the equilibrium period line. Physically, 
 the accretion induced field decay arises from the contraction of the corotation  
 radius of the magnetosphere during the accretion spin-up phase, and the evolution track 
 has little  
 chance to meet the equilibrium line if the field decay time scale is shorter 
 than the spin-up time scale at the early  stage of the accretion phase. 
 Some X-ray sources in HMXB such as Her X-1 and Vela X-1 should be very close to  
 the equilibrium period position if  the field decay really   
 exists in the accretion phase.

\begin{center}
{\bf 3.Conclusion}
\end{center}

We have presented a simple model for the evolution of the magnetic field
and spin period of accreting neutron stars. Analytic formulae for
evolution trajectories in B-P are derived. The theoretical minimum
period of the neutron star does not depend on the accretion rate
but instead depends on the total amount of accreted matter, and the stellar
 parameters including the moment of inertia, stellar mass and radius,
 which depend on the equations of state. Our model results seem
 to be supported by the observed data (White \& Zhang 1997).
 However, in this paper we have ignored the fact that the stellar
 parameters, i.e. $M$, $R$ and $I$, are all time dependent.
 The exact evolution curves must take this factor into account.
On the other hand, the minimum period should not depend on the details
of the evolution trajectories instead it only depends on the final
values of the stellar
parameters which are equations of state dependent. Observing the
minimum period in LMXB may provide useful constraints on the
equation of state for high density matter.
For $\Delta M < 1M_{\odot}$, the minimum period is longer than
1.7ms for a wide range of realistic equations of state
(Wiringa, Fiks \& Fabrocini 1988).

We thank A.Potekhin for his useful comments and P.K.MacKeown
for a critical reading of our manuscript. This research program
is partially supported by a RGC grant of the Hong Kong
Government, a Croucher Foundation Senior Research Fellowship and a NSF grant 
of PRC. 

\newpage

\begin{center}
{\bf REFERENCES}
\end{center}

\baselineskip=5mm

\noindent

\begin{description}

\item  Bailes, M. et al  1994, ApJL,425, 41

\item  Bell, J.F. et al  1997, MNRAS,286, 463

\item  Bhattacharya, D. \& van den Heuvel, E.P.J. 1991, Phys. Rep., 203,1

\item  Bhattacharya, D. \& Srinivasan, G. 1995, in X-ray Binaries, eds. 
       W.H.G. Lewin, J. van Paradijs \& E.P.J. van den Heuvel,(Cambridge
       University Press). 

\item  Bisnovatyi-Kogan, G. \& Komberg, B.1974, Soviet Astron,18,217

\item  Camilo, F. et al  1996, APJ, 469, 819    

\item  Chen, K. \& Ruderman, M.A., 1993, ApJ, 402,264

\item  Cheng, K.S.\& Dai, Z.G.1997, ApJ, 476, L39

\item  Cheng, K.S.\& Zhang, C.M. 1998, A\&A, 337, 441

\item  Chanmugam, G. 1992, Ann. Rev. Astron. Astrophys., 30,143

\item  Corbet, et al  1995,APJ,443,786    

\item  Ding, K.Y.,Cheng, K.S.\& Chau,H.F.1993,ApJ,408,167

\item  Ghosh, P. \& Lamb, F.K., 1979, ApJ, 234, 296

\item  Ghosh, P. \& Lamb, F.K., 1991 in "Neutron Stars :
Theory and Observation" eds
J. Ventura \& D. Pines (Kluwer:Netherlands), p.363

\item  Lorimer, D.R., et al 1996, MNRAS, 283, 1383    

\item Makishima, K. 1992, in "The structure and evolution of neutron stars", eds
D. Pines, R. Tamagaki \& S. Tsuruta (Addison-Wesley), p.86

\item  Phinney, E.S. \& Kulkarni, S.R. 1994, Ann. Rev. Astron. Astrophys.,
32, 591.

\item  Romani, G.W. 1990,Nature,347,741

\item  Ruderman, M. 1991a,ApJ,366,261

\item  Ruderman, M. 1991b,ApJ,382,576

\item  Ruderman, M. 1991c,ApJ,382,587

\item  Ruderman, M., Zhu, T. \& Chen, K. 1998,ApJ,492, 267

\item  Shibazaki, N., Murakami, T., Shaham, J. \& Nomoto, K. 1989,
Nature,342,656

\item  Taam, R.E. \& van den Heuvel, E.P.J. 1986, ApJ, 305, 235

\item Taylor, J.H., Manchester, R.N. \& Lyne, A.G. 1993, APJS,88,529

\item  Urpin, V. \& Geppert, U. 1995, MNRAS,275,1117

\item  Urpin, V. \& Konenkov, D. 1997,MNRAS,284,741

\item  van den Heuvel, E.P.J., van Paradijs, J.A.
\& Taam, R.E.1986,Nature,322,153

\item  van den Heuvel, E.P.J. ,Bizaraki, O .1995a,A\&A,297,L41

\item  van den Heuvel, E.P.J. ,Bizaraki, O. 1995b,in The Lives of the Neutron
Stars, eds. M.A. Alpar, $\ddot{{\rm U}}$. Kilzilo$\check{{\rm g}}$lu and J.
van Paradijs, (Kluwer Academic Publishers, Dordrecht).

\item  White, N. \& Zhang, W. 1997, APJ, 490, L87

\item   Wiringa, R.B., Fiks, V.  \& Fabrocini, A., 1988, Phys. Rev. C38, 1010

\item  Zhang, C.M., Wu, X.J. \& Yang, G.C.1994,A\&A,283,889
\item  Zhang, C.M., 1998, A\&A, 330, 195   
\item  Zhang, C.M.,  Cheng, K.S., \& Zhang, L. 1997, in "21st Centuary 
Chinese Astronomy Conference", eds. K.S.Cheng \& S.Chan,
(World Scientific, Singapore), p301
 
\end{description}

\newpage
\begin{figure}
\vskip -2cm
\psfig{file=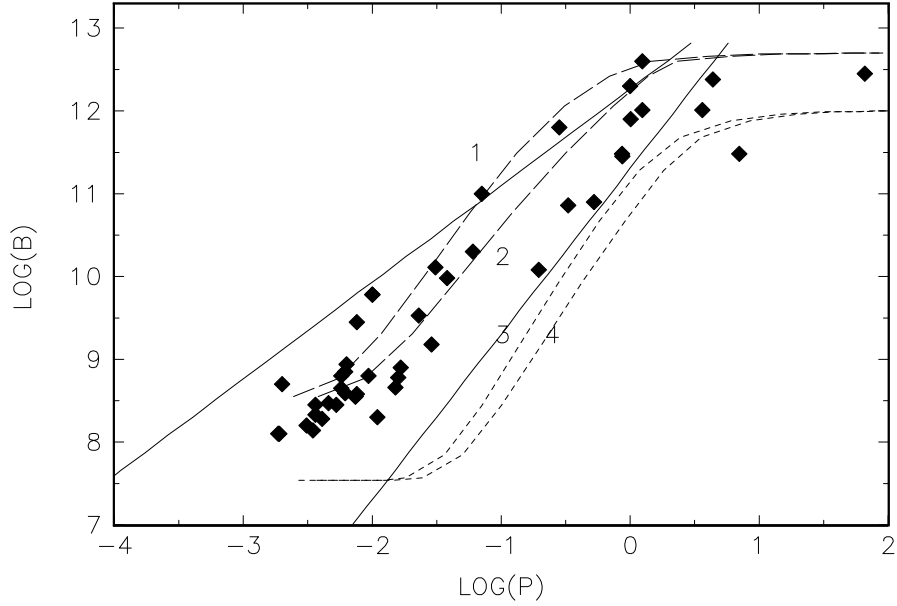,width=0.945\textwidth,angle=270}
\caption{Magnetic field vs. spin period diagram.
Solid diamonds are observed data summarized in table 1.
Initial spin
period is chosen to be 100 s. The initial magnetic fields
$B_0 =5\times 10^{12}$ G and $B_0 =1\times 10^{12}$ G are used, and
the luminosity is varied from the Eddington limited
luminosity $L_{38}$=1 to $L_{36}$=1. The upper heavy solid
line is the spin-up line (equilibrium period line) given in equation (11)
and the lower solid  line is
the death line defined  as $B_{12}/P^2$ = 0.2. Curve 1
and  curve 3 are calculated from equation (6) with luminosity $L_{38}=1$
and $L_{36}=1$ respectively.  Curve 2 and curve 4 are our
numerical solutions for
equations (1)-(5) with luminosity $L_{38}$=1. and $L_{36}$=1. respectively.
In all model evolution curves, we have used M = 1$M_{\odot}$ and R = 10$^6$cm.}
\end{figure}

\end{document}